\documentclass{ws-rv975x65}
\usepackage{subfigure}     
\usepackage{ws-rv-van}     
\usepackage{harpoon}
\makeindex

\usepackage{chngcntr}
\counterwithout{chapter}{chapter}
\counterwithout{section}{section}
\counterwithout{equation}{section}
\counterwithout{figure}{section}
\counterwithout{table}{section}

\begin{document}

\begin{center}
{\Large\bf Neutrino Masses and Flavor Oscillations}
\end{center}
\vspace{0.5cm}

\author[Y.F. Wang $\&$ Z.Z. Xing]{{\bf Yifang Wang}\footnote{E-mail: yfwang@ihep.ac.cn},
~ {\bf Zhi-zhong Xing}\footnote{E-mail: xingzz@ihep.ac.cn}}

\address{Institute of High Energy Physics, Chinese Academy of Sciences, \\
P.O. Box 918, Beijing 100049, China}

\begin{center}
{\bf Abstract}
\end{center}

\begin{abstract}
This essay is intended to provide a brief description of the
peculiar properties of neutrinos within and beyond the standard
theory of weak interactions. The focus is on the flavor
oscillations of massive neutrinos, from which one has achieved some
striking knowledge about their mass spectrum and flavor
mixing pattern. The experimental prospects towards probing the
absolute neutrino mass scale, possible Majorana nature and
CP-violating effects will also be addressed.
\end{abstract}
\markright{Neutrino Masses and Flavor Oscillations}
\body

\vspace{1cm}

\section{Neutrinos and Their Sources}

\subsection{From Pauli's hypothesis to the discoveries of neutrinos}

Soon after Henri Becquerel discovered the radioactivity of uranium in
1896 \cite{Becquerel},
many nuclear physicists started to pay attention to the
beta decays $(A, Z) \to (A, Z+1) + e^-$, in which the energy
spectrum of electrons was expected to be {\it discrete} thanks to
the laws of energy and momentum conservations. However, James
Chadwick observed a {\it continuous} electron energy spectrum of the
beta decay in 1914 \cite{Chadwick1}, and such a result was firmly
confirmed by Charles Ellis and his colleagues in the 1920s
\cite{Ellis}. At that time there were two different ideas to resolve
this ``new physics" phenomenon (i.e., the discrepancy between {\it
observed} and {\it expected} energy spectra of electrons): one was
to give up the energy conservation law and the other was to add in a
new particle. Niels Bohr was the representative of the former idea,
which turned out to be wrong. Wolfgang Pauli conjectured that an
unobservable, light, spin-1/2 and neutral particle --- known as the
electron antineutrino later --- appeared in the beta decay and
carried away some energy and momentum, and thus the energy spectrum
of electrons in the process $(A, Z) \to (A, Z+1) + e^- +
\overline{\nu}^{}_e$ was continuous. Pauli first put forward the
concept of neutrinos in his famous letter to the ``Dear radioactive
ladies and gentlemen" who had gathered in T$\rm\ddot{u}$bingen on 4
December 1930 \cite{Pauli}. Three years later he gave a talk on his
neutrino hypothesis in the renowned Solvay Conference, where Enrico
Fermi was in the audience and took this hypothesis seriously.
In the end of 1933,
Fermi published his most important theoretical work --- an effective
theory of the beta decay \cite{Fermi}, which is actually a
low-energy version of today's standard picture of weak
charged-current interactions. Fermi's seminal work made it possible
to calculate the reaction rates of nucleons and electrons (or
positrons) interacting with neutrinos (or antineutrinos).

In 1936, Hans Bethe pointed out that an inverse beta decay mode of
the type $\overline{\nu}^{}_e + p \to n + e^+$ (or more general,
$\overline{\nu}^{}_e + (A, Z) \to (A, Z-1) + e^+$) could be
a possible way to verify the existence of electron antineutrinos
produced from either fission bombs or fission reactors \cite{Bethe0}.
This preliminary idea was elaborated by Bruno Pontecorvo in
1946 \cite{Pon46}, and it became feasible with the development of
the liquid scintillation counting techniques in the 1950s.
Although the incident $\overline{\nu}^{}_e$ is invisible, it can
trigger the inverse beta decay where the emitted positron
annihilates with an electron and the daughter nucleus is captured in
the detector. Both events are observable because they emit gamma
rays, and the corresponding flashes in the liquid scintillator
are separated by some
microseconds. Frederick Reines and Clyde Cowan did the first reactor
antineutrino experiment and obtained a positive result in 1956
\cite{RC}, and they reported a new result consistent with the
parity-violating theory of weak interactions in 1960.
The Nobel Prize finally came to Reines in 1995, when Cowan had
passed away 21 years before.

The discovery of electron antineutrinos motivated Pontecorvo
to speculate on the possibility of lepton number violation and
neutrino-antineutrino transitions in 1957 \cite{Pontecorvo}. His
argument was actually based on a striking conjecture made by Ettore
Majorana in 1937: a massive neutrino could be its own antiparticle
\cite{Majorana}.

In 1962, the muon neutrino --- a sister of the electron neutrino ---
was discovered by Leon Lederman, Melvin Schwartz and Jack Steinberger
in an accelerator-based experiment \cite{Danby}. This discovery,
which immediately motivated Ziro Maki, Masami Nakagawa and Shoichi
Sakata to conjecture the $\nu^{}_e \leftrightarrow \nu^{}_\mu$ conversion
\cite{MNS}, was also recognized by the Nobel Prize in 1988. The tau
neutrino, another sister of the electron neutrino, was finally
observed at the Fermilab in the end of 2000 \cite{Kodama}. Within
the standard model the complete lepton family consists of three
charged members ($e$, $\mu$, $\tau$) and three neutral members
($\nu^{}_e$, $\nu^{}_\mu$, $\nu^{}_\tau$), and their corresponding
antiparticles.

\subsection{Where do neutrinos come from?}

Neutrinos and antineutrinos may originate from many
physical and astrophysical processes via weak interactions.
Fig. 1 illustrates some typical examples of neutrino or antineutrino
sources in the Universe.
\begin{figure}[t]
\centerline{\includegraphics[width=11.5cm]{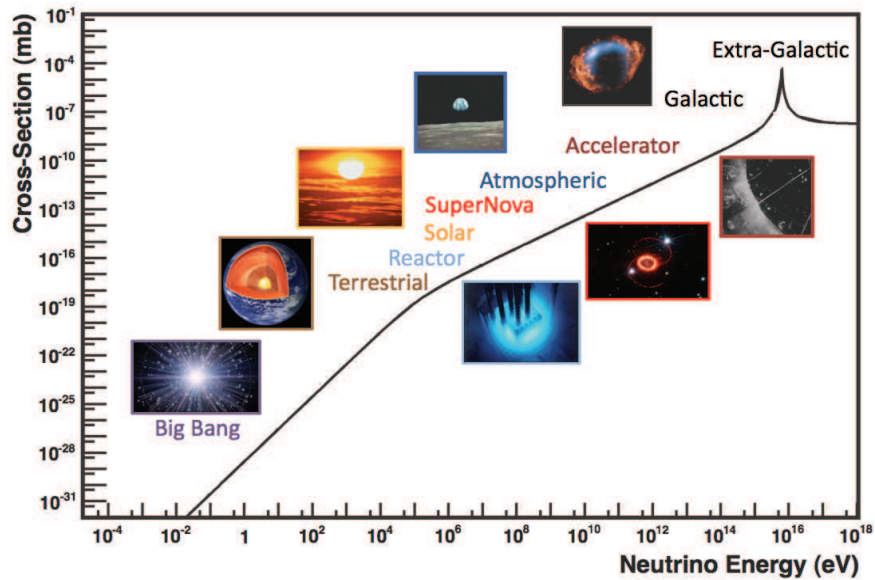}} \caption{
Some representative
sources of neutrinos and (or) antineutrinos and their corresponding
energies \cite{Source}. The cross sections of
$\overline{\nu}^{}_e + e^- \to \overline{\nu}^{}_e + e^-$
scattering associated with different sources are also shown
for comparison, where the peak around $6.3$ PeV
is related to the Glashow resonance \cite{Glashow}.}
\end{figure}

{\it Example (1)}: Neutrinos and antineutrinos from the Big Bang.
The standard cosmology predicts the existence of a cosmic neutrino
(or antineutrino) background in the Universe.
Today such {\it relic} neutrinos and antineutrinos should have
an overall number density around $330 ~{\rm cm}^{-3}$, but their
temperature is so low (only about $1.9$ K, or roughly $1.6 \times
10^{-4}$ eV) that there is no way to detect them. In the long run it
might be possible to capture the relic electron neutrinos on some
beta-decaying nuclei \cite{W}, as the PTOLEMY project is trying
\cite{P}.

{\it Example (2)}: Electron antineutrinos from the Earth. Since its
birth, the Earth's interior has kept a number of radioactive nuclei (e.g.,
$^{40}{\rm K}$, $^{238}{\rm U}$ and $^{232}{\rm Th}$). That is why
numerous electron antineutrinos can be produced from terrestrial
``natural radioactivity" (i.e., the beta decays), at a rate of several
millions per square centimeter per second. So far such interesting
geo-$\overline{\nu}^{}_e$ events have been observed at the $3\sigma$
level in the KamLAND \cite{GEO1} and Borexino \cite{GEO2} experiments.

{\it Example (3)}: Electron neutrinos from the Sun. Solar electron
neutrinos come along with a number of thermonuclear fusion reactions
inside the Sun. One may understand why the Sun shines with the help
of $4 p \to ~^4{\rm He} + 2 e^+ + 2 \nu^{}_e + 26.7 ~{\rm MeV}$:
about $98\%$ of the energy radiates in the form of light and only
$2\%$ of the energy is taken away by neutrinos \cite{Bethe}. The
only way to verify such a picture on the Earth is to detect the
electron neutrinos emitted from the core of the Sun. In 1968 solar
neutrinos were first observed by Raymond Davis in his radiochemical
experiment (see section 4.1 for a more detailed description)
\cite{Davis}.

{\it Example (4)}: Neutrinos and antineutrinos from supernovae. The
explosion of a supernova may release the gravitational binding energy of
${\cal O}(10^{53})$ erg in the form of neutrinos and antineutrinos
\cite{Bethe2}. On 23 February 1987 the $\nu^{}_e$ and
$\overline{\nu}^{}_e$ events from the Supernova 1987A explosion were
observed by the Kamiokande-II \cite{Koshiba}, IMB \cite{IMB} and
Baksan \cite{Baksan} detectors. This observation was a great
milestone in neutrino astronomy. Davis and Masatoshi Koshiba received the
Nobel Prize in 2002 for their pioneering detections of solar and
supernova neutrinos, respectively.

{\it Example (5)}: Neutrinos and  antineutrinos from the Earth's
atmosphere. When a cosmic ray (which is mainly composed of
high-energy protons
coming from somewhere in the galactic or extragalactic space)
penetrates the atmosphere around the Earth, it may interact with the
ambient nuclei and generate a particle shower containing charged
pions and muons. The decays of $\pi^{\pm}$ and $\mu^{\pm}$ can
therefore produce atmospheric $\nu^{}_\mu$, $\overline{\nu}^{}_\mu$,
$\nu^{}_e$ and $\overline{\nu}^{}_e$ events, which have been
observed in several experiments \cite{PDG}. In particular, the
phenomenon of atmospheric neutrino oscillations was firmly
established by the Super-Kamiokande (SK) Collaboration in 1998 \cite{SK2}.

{\it Example (6)}: Ultrahigh-energy (UHE) cosmic neutrinos and
antineutrinos from distant astrophysical sources, including the
expected active galactic nuclei, gamma ray bursts, supernova
remnants and the Greisen-Zatsepin-Kuzmin cutoff of cosmic rays
\cite{XZ}. The UHE $\nu^{}_\mu$, $\overline{\nu}^{}_\mu$, $\nu^{}_e$
and $\overline{\nu}^{}_e$ events can be produced from UHE $p\gamma$
or $pp$ collisions via $\pi^{\pm}$ and $\mu^{\pm}$ decays, and thus
they may serve as a unique cosmic messenger and provide us with
useful information about the cosmos that cannot be extracted from
the measurements of cosmic rays and gamma rays. So far the IceCube
detector at the South Pole has observed 37 extraterrestrial neutrino
candidate events with deposited energies ranging from 30 TeV to 2
PeV \cite{IC}. Among them, the three PeV events represent the
highest-energy neutrino interactions ever observed, but their
astrophysical origin remains mysterious.

Of course, neutrinos and (or) antineutrinos can also be produced
from some man-made facilities, especially the nuclear reactors and
particle accelerators. They also play a crucial role in discovering
neutrinos, observing flavor oscillations and measuring fundamental
parameters, as one will see in sections 3---5.

\section{Weak Interactions of Neutrinos in the Standard Theory}

As an important part of the matter content in the standard
electroweak model based on the $SU(2)^{}_{\rm L} \times U(1)^{}_{\rm
Y}$ gauge group, neutrinos are assumed to be the {\it massless} Weyl
particles. Hence only the left-handed neutrinos and right-handed
antineutrinos exist, and they take part in weak charged- and
neutral-current interactions via
\begin{eqnarray}
-{\cal L}^{}_{\rm cc} & = & \frac{g}{2\sqrt{2}} \sum_\alpha \left[
\overline{\alpha} \ \gamma^\mu \left(1 - \gamma^{}_5\right)
\nu^{}_\alpha W^-_\mu + {\rm h.c.} \right] \; ,
\nonumber \\
-{\cal L}^{}_{\rm nc} & = & \frac{g}{4\cos\theta^{}_{\rm w}}
\sum_\alpha \left[
\overline{\nu^{}_\alpha} \ \gamma^\mu \left(1 - \gamma^{}_5\right)
\nu^{}_\alpha \right] Z^{}_\mu \; ,
\end{eqnarray}
where $\alpha = e, \mu, \tau$. Eq. (1) allows one to calculate the
cross sections of neutrino-electron, neutrino-neutrino and
neutrino-nucleon scattering processes \cite{XZ}. Note that the
reactions $\nu^{}_e + e^- \to \nu^{}_e + e^-$ and
$\overline{\nu}^{}_e + e^- \to \overline{\nu}^{}_e + e^-$ can happen
via both charged- and neutral-current interactions, but $\nu^{}_\mu
+ e^- \to \nu^{}_\mu + e^-$ (or $\nu^{}_\tau + e^- \to \nu^{}_\tau +
e^-$) and $\overline{\nu}^{}_\mu + e^- \to \overline{\nu}^{}_\mu +
e^-$ (or $\overline{\nu}^{}_\tau + e^- \to \overline{\nu}^{}_\tau +
e^-$) can only occur via the neutral-current interactions. That is
why the behavior of neutrino flavor conversion in a dense medium may
be modified by the coherent forward $\nu^{}_e e^-$ or
$\overline{\nu}^{}_e e^-$ scattering. This effect is referred to as
the Wolfenstein-Mikheyev-Smirnov (MSW) matter effect \cite{MSW}.

The simplest quasi-elastic neutrino-nucleon scattering processes are
the inverse beta decays $\overline{\nu}^{}_e + p \to e^+ + n$ and
$\nu^{}_e + n \to e^- + p$, which take place via the charged-current
weak interactions. Their cross sections can be approximately
expressed as $\sigma \left(\overline{\nu}^{}_e p\right) = \sigma
\left(\nu^{}_e n\right) \simeq 9.1 \times 10^{-44}
\left(E^{}_\nu/{\rm MeV}\right)^2 {\rm cm}^2$. In comparison, the
elastic neutrino-nucleon scattering reaction $\nu^{}_\alpha + N \to
\nu^{}_\alpha + N$ (for $\alpha = e, \mu, \tau$) is mediated by the
neutral-current weak interactions.

Historically, the existence of weak neutral currents was first
established in the Gargamelle bubble chamber at CERN in 1973
\cite{NC}. This experiment, which observed the highly expected
events of $\nu^{}_\mu + N \to \nu^{}_\mu + {\rm hadrons}$ and
$\overline{\nu}^{}_\mu + N \to \overline{\nu}^{}_\mu + {\rm
hadrons}$, crowned the long-range neutrino program initiated by CERN
at that time and brought CERN a leading role in the field of high
energy physics. It also provided an unprecedentedly strong support
to the standard electroweak model formulated by Sheldon Glashow,
Steven Weinberg and Abdus Salam in the 1960s \cite{GWS}. These three
theorists received the Nobel Prize in 1979 for their contributions to
the electroweak theory and especially for their prediction of the
weak neutral current. Four years later, the three mediators of the
weak force (i.e., the $W^\pm$ and $Z^0$ bosons) were finally
discovered by Carlo Rubbia and his colleagues at CERN \cite{Rubbia}.

The standard model was thoroughly tested in the 1990s with the help
of the Large Electron-Positron Collider (LEP) running on the $Z^0$
resonance at CERN. In particular, the number of neutrino species was
determined to be $N^{}_\nu = 2.984 \pm 0.008$ via the decay
$Z^0 \to \nu^{}_\alpha + \overline{\nu}^{}_\alpha$ \cite{PDG}. Such
a result is consistent very well with 3 as required in the theory.
Extra light neutrino species are not impossible, but they must be
``sterile" --- in the sense that they do not directly take part in
the standard weak interactions, and hence their existence is not
subject to the LEP measurement.

Note that the structure of the standard model itself is too
economical to allow the neutrinos to be massive. On the one hand,
the particle content of the model is so limited that there are
neither right-handed neutrinos nor
any Higgs triplets. Hence a normal Dirac neutrino mass term is not
allowed, nor a gauge-invariant Majorana mass term. On the other
hand, the model is a renormalizable quantum field theory. The
renormalizability implies that an effective dimension-5 operator,
which can give each neutrino a Majorana mass, is also forbidden.

\section{Neutrino Masses, Flavor Mixing and Oscillations}

\subsection{Massive neutrinos and their electromagnetic properties}

There are several ways to slightly extend the standard theory such
that the neutrinos can acquire their masses with little influence on
the great success of the theory itself \cite{Xing09}.
Here let us take two typical examples for illustration.

(1) If the renormalizability of the standard theory is relaxed, then
the lowest-dimension operator that violates lepton number and
generates neutrino masses must be the unique dimension-5 Weinberg
operator $HH \ell\ell/\Lambda$, where $\Lambda$ denotes the cut-off
energy scale in such an effective field theory, $H$ and $\ell$ are
the Higgs and lepton doublets, respectively \cite{Weinberg}. After
spontaneous gauge symmetry breaking, this operator yields the neutrino
masses $m^{}_i \sim \langle H\rangle^2/\Lambda$ (for $i=1,2,3$),
which can be sufficiently small
($\lesssim 1$ eV) provided $\Lambda \gtrsim 10^{13}$ GeV and
$\langle H\rangle \sim 10^2$ GeV. In this sense the study of
neutrino mass generation can serve as a striking low-energy window onto new
physics at superhigh energy scales.

(2) If two or more heavy right-handed neutrinos are added into the
standard theory and lepton number is violated by their Majorana mass
term, then the Lagrangian responsible for neutrino masses can be
written as
\begin{eqnarray}
-{\cal L}^{}_{\rm mass} = \overline{\ell^{}_{\rm L}} Y^{}_\nu
\tilde{H} N^{}_{\rm R} + \frac{1}{2} \overline{N^c_{\rm R}} M^{}_{\rm R}
N^{}_{\rm R} + {\rm h.c.} \; ,
\end{eqnarray}
in which the first term stands for the neutrino Yukawa interactions,
and the second term is lepton-number-violating. After the
$SU(2)^{}_{\rm L} \times U(1)^{}_{\rm Y}$ gauge symmetry is
spontaneously broken to $U(1)^{}_{\rm em}$, one is left with the
effective Majorana neutrino mass matrix $M^{}_\nu \simeq - \langle
H\rangle^2 Y^{}_\nu M^{-1}_{\rm R} Y^T_\nu$, which is often referred
to as the canonical {\it seesaw} formula \cite{SS}. Because
$N^{}_{\rm R}$ is the $SU(2)^{}_{\rm L}$ singlet, the mass scale of
$M^{}_{\rm R}$ can be greatly higher than the electroweak scale
$\langle H\rangle$. Hence the mass scale of $M^{}_\nu$ is highly
suppressed, providing a natural explanation of the smallness of
neutrino masses.

Instead of introducing the heavy right-handed neutrinos, one may
also introduce a Higgs triplet or a few triplet fermions into the
standard theory so as to explain why the three active neutrinos
should have naturally small masses \cite{XZ}. Such seesaw mechanisms
essentially have the same spirit, which attributes the smallness of
neutrino masses to the largeness of new degrees of freedom.
Furthermore, they require massive neutrinos to be the Majorana
particles and thus allow some lepton-number-violating processes to
happen.

It is worth pointing out that a pure Dirac neutrino mass term,
originating from the neutrino Yukawa interactions on the right-hand
side of Eq. (2), is less convincing and less interesting from a
theoretical point of view. The reason for this argument is two-fold:
(a) such a scenario cannot explain why the neutrino masses are so
small as compared with the charged lepton masses; (b) given
$N^{}_{\rm R}$, the lepton-number-violating term $\overline{N^c_{\rm
R}} M^{}_{\rm R} N^{}_{\rm R}$ should not be absent because it is
not forbidden by gauge symmetry and Lorentz invariance. If massive
neutrinos really have the Majorana nature, they can trigger the
neutrinoless double-beta ($0\nu\beta\beta$) decays and some other
lepton-number-violating processes. In particular, they are likely to
have something to do with the observed asymmetry of matter and
antimatter in the Universe via the seesaw and leptogenesis \cite{FY}
mechanisms. Hence the phenomenology of Majorana neutrinos is much
richer and more interesting than that of Dirac neutrinos.

Although a massive neutrino does not possess any electric charge, it
can have electromagnetic interactions via quantum loops \cite{DM}.
Now that Dirac and Majorana neutrinos couple to the photon in
different ways, their corresponding electromagnetic form factors
must be different. Given the standard weak interactions, one finds
that a massive Dirac neutrino has no electric dipole moment and its
magnetic dipole moment is finite but extremely small: $\mu^{}_\nu
\sim 3 \times 10^{-20} \left(m^{}_\nu/0.1 ~{\rm eV}\right)
\mu^{}_{\rm B}$ with $\mu^{}_{\rm B}$ being the Bohr magneton. In
contrast, a massive Majorana neutrino has neither electric nor
magnetic dipole moments, simply because its antiparticle is just
itself.

But both Dirac and Majorana neutrinos can have the {\it transition}
dipole moments (i.e., from one mass eigenstate to another mass
eigenstate), which may result in neutrino decays, neutrino-electron
scattering, neutrino interactions with external magnetic fields, etc
\cite{Giunti1}. In a realistic neutrino-electron scattering
experiment, what can be constrained is actually an effective
transition dipole moment $\mu^{}_{\rm eff}$ consisting of both
electric and magnetic components. Hence it is practically impossible
to distinguish between Dirac and Majorana neutrinos in such
measurements. Current experimental upper bounds on $\mu^{}_{\rm
eff}$ are at the level of $10^{-11} \mu^{}_{\rm B}$ \cite{Giunti1},
far above the afore-mentioned theoretical expectation $\mu^{}_\nu
\sim 10^{-20} \mu^{}_{\rm B}$.

\subsection{Lepton flavor mixing and neutrino oscillations}

In the basis where the flavor eigenstates of three charged leptons
are identified with their mass eigenstates, one may diagonalize the
Majorana neutrino mass matrix $M^{}_\nu$ by means of a unitary
transformation. Then the leptonic charged-current interactions in
Eq. (1) can be reexpressed in terms of the mass eigenstates:
\begin{eqnarray}
-{\cal L}^{}_{\rm cc} = \frac{g}{\sqrt{2}} \ \overline{\left( e ~
\mu ~ \tau\right)^{}_{\rm L}} \ \gamma^\mu \ U \left(
\begin{matrix} \nu^{}_1 \cr \nu^{}_2 \cr \nu^{}_3
\end{matrix} \right)^{}_{\rm L} W^-_\mu + {\rm h.c.} \; ,
\end{eqnarray}
where the $3\times 3$ unitary matrix $U$ describes the strength of
lepton flavor mixing and can be parameterized by using three
rotation angles and three CP-violating phases:
\begin{eqnarray}
U = \left(
\begin{matrix} c^{}_{12} c^{}_{13} & s^{}_{12} c^{}_{13} & s^{}_{13}
e^{-{\rm i} \delta} \cr -s^{}_{12} c^{}_{23} - c^{}_{12} s^{}_{13}
s^{}_{23} e^{{\rm i} \delta} & c^{}_{12} c^{}_{23} - s^{}_{12}
s^{}_{13} s^{}_{23} e^{{\rm i} \delta} & c^{}_{13} s^{}_{23} \cr
s^{}_{12} s^{}_{23} - c^{}_{12} s^{}_{13} c^{}_{23} e^{{\rm i}
\delta} & ~ -c^{}_{12} s^{}_{23} - s^{}_{12} s^{}_{13} c^{}_{23}
e^{{\rm i} \delta} ~ & c^{}_{13} c^{}_{23} \cr
\end{matrix} \right) P^{}_\nu \; ,
\end{eqnarray}
where $c^{}_{ij} \equiv \cos\theta^{}_{ij}$, $s^{}_{ij} \equiv
\sin\theta^{}_{ij}$ (for $ij = 12, 13, 23$), $\delta$ is referred to
as the Dirac CP-violating phase, and $P^{}_\nu = {\rm
Diag}\left\{e^{{\rm i}\rho}, e^{{\rm i}\sigma}, 1\right\}$ contains
two extra phase parameters of the Majorana nature. The matrix $U$ is
often called the Pontecorvo-Maki-Nakagawa-Sakata (PMNS) matrix, and
its unitarity has been tested at the percent level \cite{Antusch}
\footnote{Note that whether $U$ is unitary or not depends on the
mechanism of neutrino mass generation. In the canonical seesaw
mechanism \cite{SS}, for instance, the mixing between light and
heavy Majorana neutrinos may lead to tiny unitarity-violating
effects for the PMNS matrix $U$ itself.}.

Eq. (3) tells us that a $\nu^{}_\alpha$ neutrino can be produced
from the $W^+ + \alpha^- \to \nu^{}_\alpha$ interaction, and a
$\nu^{}_\beta$ neutrino can be detected through the $\nu^{}_\beta
+ W^- \to \beta^-$ interaction (for $\alpha, \beta = e, \mu, \tau$).
The $\nu^{}_\alpha \to \nu^{}_\beta$ oscillation may happen if
the $\nu^{}_i$ beam with energy $E \gg m^{}_i$ travels a proper
distance $L$ in vacuum. The probability of such a flavor oscillation
is given by \cite{XZ}
\begin{eqnarray}
P(\nu^{}_\alpha \to \nu^{}_\beta) & = & \delta^{}_{\alpha\beta} - 4
\sum_{i<j} \left({\rm Re} \lozenge^{ij}_{\alpha\beta}
\sin^2\Delta^{}_{ji} \right)
+ \ 8 {\rm Im} \lozenge^{ij}_{\alpha\beta}
\prod_{i<j} \sin\Delta^{}_{ji} \; ,
\end{eqnarray}
in which $\Delta^{}_{ji} \equiv \Delta m^2_{ji} L/\left(4 E\right)$
and $\lozenge^{ij}_{\alpha\beta} \equiv U^{}_{\alpha i} U^{}_{\beta
j} U^*_{\alpha j} U^*_{\beta i}$ (for $i,j = 1,2,3$ and $\alpha,
\beta = e, \mu, \tau$). The probability of the
$\overline{\nu}^{}_\alpha \to \overline{\nu}^{}_\beta$ oscillation
can easily be read off from Eq. (5) by making the replacement $U \to
U^*$. There are two types of neutrino oscillation experiments: the
``appearance" one ($\alpha \neq \beta$) and the ``disappearance" one
($\alpha = \beta$). Both solar neutrino oscillations ($\nu^{}_e \to
\nu^{}_e$) and reactor antineutrino oscillations
($\overline{\nu}^{}_e \to \overline{\nu}^{}_e$) are of the
disappearance type. The atmospheric muon-neutrino (or
muon-antineutrino) oscillations essentially belong to the
disappearance type, and the accelerator neutrino oscillations can be
of either type.

At this point let us explain why it is extremely difficult to do a
realistic neutrino-antineutrino oscillation experiment. We consider
an $\overline{\nu}^{}_\alpha$ beam produced from the standard
charged-current interactions $\alpha^+ + W^- \to
\overline{\nu}^{}_\alpha$. After traveling a distance $L$ this beam
will be detected at a detector through the standard charged-current
interactions $\nu^{}_\beta \to \beta^- + W^+$. Different from the
normal $\nu^{}_\alpha \to \nu^{}_\beta$ or $\overline{\nu}^{}_\alpha
\to \overline{\nu}^{}_\beta$ oscillations, the
$\overline{\nu}^{}_\alpha \to \nu^{}_\beta$ oscillation involves a
suppression factor $m^{}_i/E$ in its amplitude. This factor reflects
the fact that the incoming $\alpha^+$ leads to an antineutrino
$\overline{\nu}^{}_\alpha$ in a dominantly right-handed helicity
state, whereas the standard charged-current interactions that
produce the outgoing $\beta^-$ would prefer the incident neutrino
$\nu^{}_\beta$ being in a left-handed state \cite{S}. Because of $m^{}_i
\lesssim 1$ eV and $E \gtrsim 1$ MeV in a realistic experiment, this
helicity suppression factor (i.e., $m^{}_i/E \lesssim 10^{-6}$)
makes it impossible to observe the phenomenon of
neutrino-antineutrino oscillations.

\section{Observations of Neutrino Oscillations}

\subsection{Solar neutrino oscillations}

In 1946 Pontecorvo put forward a radiochemical technique which can
be used to measure solar electron neutrinos via the reaction
$^{37}{\rm Cl} + \nu^{}_e \to ~^{37}{\rm Ar} + e^-$ \cite{Pon46}.
The incident neutrino's energy threshold for this reaction to happen
is $0.814$ MeV, low enough to make it sensitive to solar $^8{\rm B}$
neutrinos. In 1964 John Bahcall carefully calculated the solar
neutrino flux and the capture rate of $^8{\rm B}$ neutrinos,
demonstrating the experimental feasibility of Pontecorvo's idea
\cite{Bahcall}. This motivated Davis to build a $10^5$-gallon
Chlorine-Argon neutrino detector in the Homestake Gold Mine in the
middle of the 1960s. The final result of this experiment was published
in 1968 and caused a big puzzle: the measured flux of solar $^8{\rm
B}$ neutrinos was only about one third of the value predicted by the
standard solar model (SSM) \cite{Davis}. Such a deficit was later
confirmed in a number of solar neutrino experiments, including the
Homestake \cite{H}, GALLEX/GNO \cite{G}, SAGE \cite{Sage}, SK
\cite{SK} and SNO \cite{SNO} experiments. Among them, the SNO
experiment was especially crucial because it
model-independently demonstrated the flavor conversion of solar
$\nu^{}_e$ neutrinos into $\nu^{}_\mu$ and $\nu^{}_\tau$
neutrinos.
\begin{figure}[t]
\centerline{\includegraphics[width=12.5cm]{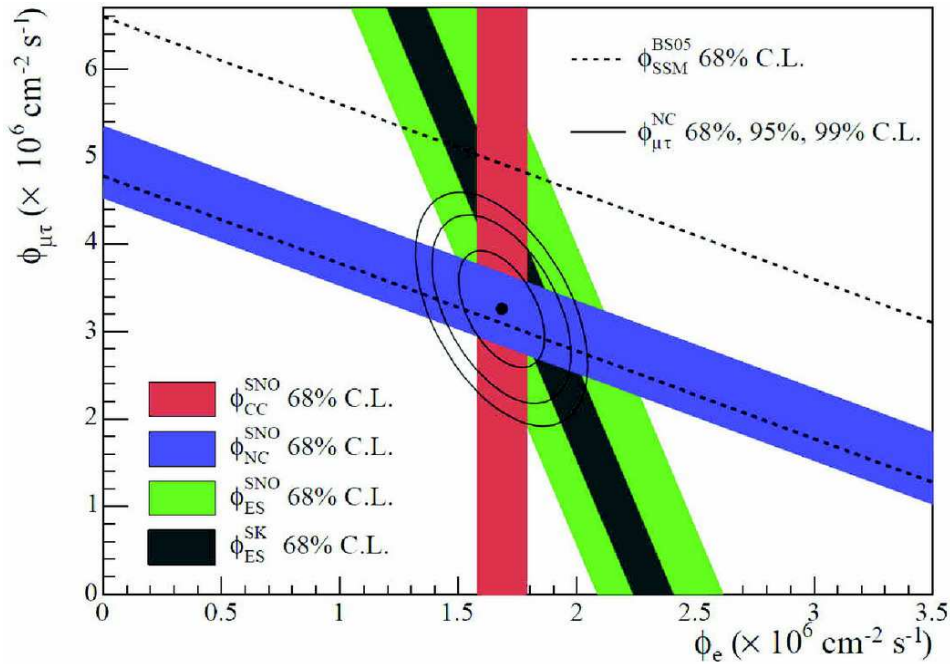}} \caption{The
$\nu^{}_\mu + \nu^{}_\tau$ flux versus the $\nu^{}_e$ flux
determined from the SNO data. The total solar $^8{\rm B}$ neutrino
flux predicted by the SSM is shown as dashed lines, parallel to the
NC measurement. The narrowed band parallel to the SNO's ES
measurement corresponds to the SK's ES result. The best-fit point is
obtained by using only the SNO data \cite{SNO2}.}
\end{figure}

Given heavy water as the target material of the SNO detector, the
solar $^8{\rm B}$ neutrinos were measured via the charged-current
(CC) reaction $\nu^{}_e + {\rm D} \to e^- + p + p$, the
neutral-current (NC) reaction $\nu^{}_\alpha + {\rm D} \to
\nu^{}_\alpha + p + n$ and the elastic-scattering process
$\nu^{}_\alpha + e^- \to \nu^{}_\alpha + e^-$ (for $\alpha = e, \mu,
\tau$) \cite{SNO}. The observed neutrino fluxes in these three
different channels are expected to satisfy $\phi^{}_{\rm CC} =
\phi^{}_e$, $\phi^{}_{\rm NC} = \phi^{}_e + \phi^{}_{\mu\tau}$ and
$\phi^{}_{\rm ES} = \phi^{}_e + 0.155 \phi^{}_{\mu \tau}$, where
$\phi^{}_{\mu\tau}$ denotes a sum of the fluxes of $\nu^{}_\mu$ and
$\nu^{}_\tau$ neutrinos. So $\phi^{}_{\rm CC} = \phi^{}_{\rm NC} =
\phi^{}_{\rm ES}$ would hold if there were no flavor conversion
(i.e., $\phi^{}_{\mu\tau} =0$). The SNO data $\phi^{}_{\rm CC} =
1.68^{+0.06}_{-0.06} ({\rm stat})^{+0.08}_{-0.09} ({\rm syst})$,
$\phi^{}_{\rm NC} = 4.94^{+0.21}_{-0.21} ({\rm
stat})^{+0.38}_{-0.34} ({\rm syst})$ and $\phi^{}_{\rm ES} =
2.35^{+0.22}_{-0.22} ({\rm stat})^{+0.15}_{-0.15} ({\rm syst})$ as
illustrated in Fig. 2 \cite{SNO2} definitely demonstrated
$\phi^{}_{\mu\tau} \neq 0$. Now we are sure that the deficit
of solar $^8{\rm B}$ neutrinos, whose typical energies
are about 6 MeV to 7 MeV, is due to $\nu^{}_e \to
\nu^{}_\mu$ and $\nu^{}_e \to \nu^{}_\tau$ oscillations
modified by significant MSW matter effects
in the Sun. A careful analysis shows that the observed survival
probability of solar $^8{\rm B}$ neutrino oscillations can
approximate to $P(\nu^{}_e \to \nu^{}_e) \simeq \sin^2\theta^{}_{12}
\simeq 0.32$ \cite{Kayser}, leading us to $\theta^{}_{12} \simeq
34^\circ$.

Moreover, the Borexino experiment has accomplished a real-time
measurement of the mono-energetic solar $^7{\rm Be}$ neutrinos with
$E = 0.862$ MeV and observed a remarkable deficit corresponding to
$P(\nu^{}_e \to \nu^{}_e) =0.56 \pm 0.1$ \cite{B}. Such a result can
roughly be explained as a vacuum oscillation effect,
because the low-energy $^7{\rm Be}$ neutrino oscillation is
not very sensitive to matter effects \cite{Kayser}. In this case we
are left with the averaged survival probability $P(\nu^{}_e \to
\nu^{}_e) \simeq 1 - \sin^2 2\theta^{}_{12}/2 \simeq 0.56$ as a
reasonable approximation for solar $^7{\rm Be}$ neutrinos, and thus
obtain $\theta^{}_{12} \simeq 35^\circ$. This result is essentially
consistent with the one extracted from solar $^8{\rm B}$ neutrinos.

\subsection{Atmospheric neutrino oscillations}

The atmospheric $\nu^{}_\mu$, $\overline{\nu}^{}_\mu$, $\nu^{}_e$
and $\overline{\nu}^{}_e$ events are produced in the Earth's
atmosphere by cosmic rays, mainly via the decays $\pi^+ \to \mu^+ +
\nu^{}_\mu$ with $\mu^+ \to e^+ + \nu^{}_e + \overline{\nu}^{}_\mu$
and $\pi^- \to \mu^- + \overline{\nu}^{}_\mu$ with $\mu^- \to e^- +
\overline{\nu}^{}_e + \nu^{}_\mu$. So the ratio of
$\nu^{}_\mu$ and $\overline{\nu}^{}_\mu$ events to $\nu^{}_e$
and $\overline{\nu}^{}_e$ events is expected to be nearly
$2:1$ at low energies ($\lesssim 1$ GeV). But a smaller
ratio was observed at the Kamiokande \cite{K2} and IMB \cite{IMB2}
detectors in the late 1980s and early 1990s,
indicating a preliminary deficit of atmospheric muon
neutrinos and muon antineutrinos. If there were no neutrino
oscillation, the atmospheric neutrinos that enter and excite an
underground detector would have an almost perfect spherical
symmetry. Namely, the downward-going and upward-going neutrino
fluxes should be equal to each other, or equivalently $\Phi^{}_e
(\theta^{}_z) = \Phi^{}_e (\pi -\theta^{}_z)$ and $\Phi^{}_\mu
(\theta^{}_z) = \Phi^{}_\mu (\pi -\theta^{}_z)$ for the zenith angle
$\theta^{}_z$. In 1998 the SK Collaboration observed
an approximate up-down flux symmetry for
atmospheric $\nu^{}_e$ and $\overline{\nu}^{}_e$ events and a
significant up-down flux asymmetry for atmospheric $\nu^{}_\mu$
and $\overline{\nu}^{}_\mu$ events \cite{SK2}.
\begin{figure}[t]
\centerline{\includegraphics[width=7.6cm]{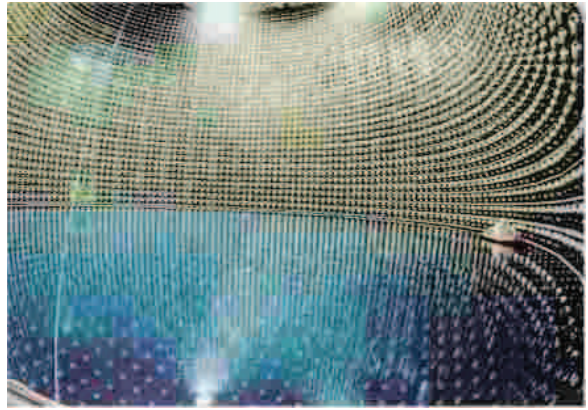}} \caption{
A brief view from
inside the SK detector's water tank during filling \cite{SK2}.}
\end{figure}

The SK detector is a $5\times 10^4$-ton tank of ultra-pure water,
located approximately 1 km underground in the Mozumi Mine in
Kamioka. As illustrated in Fig. 3, the inside surface of the tank is
lined with more than $1.1 \times10^4$ photo-multiplier tubes (PMTs).
An additional layer of water called the outer detector is also
instrumented PMTs to detect any charged particles entering the
central volume and to shield the inner detector by absorbing any
neutrons produced in the nearby rock. A neutrino interacting with
the electrons or nuclei of water can produce a charged particle that
moves faster than the speed of light in water, creating a cone of
light known as Cherenkov radiation. The Cherenkov light is projected
as a ring on the wall of the detector and recorded by the PMTs.
Hence the direction and flavor of an incident neutrino can be
identified by using the details of the ring pattern.

As shown in Fig. 4, the observed deficit of atmospheric upward-going
$\nu^{}_\mu$ and $\overline{\nu}^{}_\mu$ events at SK could
naturally be attributed to $\nu^{}_\mu \to \nu^{}_\tau$ and
$\overline{\nu}^{}_\mu \to \overline{\nu}^{}_\tau$ oscillations,
because the detector itself was insensitive to $\nu^{}_\tau$ and
$\overline{\nu}^{}_\tau$ events. This was actually the first {\it
model-independent} evidence for neutrino oscillations, and it marked
the threshold of a new era in particle physics. Since 1998 a number
of breakthroughs have been made in experimental neutrino physics.
\begin{figure}[t]
\centerline{\includegraphics[width=8.5cm]{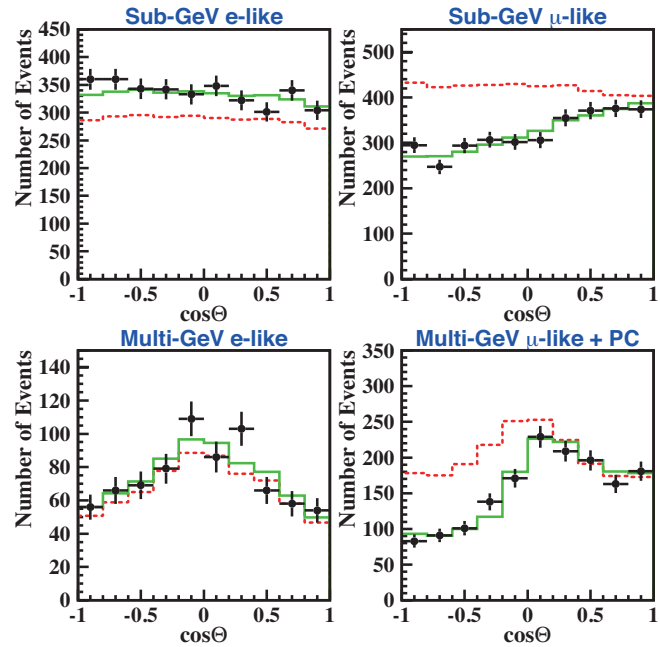}} \caption{The SK
zenith-angle distributions for fully contained 1-ring $e$-like and
$\mu$-like events with visible energy $< 1.33$ GeV (sub-GeV) and $>
1.33$ GeV (multi-GeV). For multi-GeV $\mu$-like events, a combined
distribution with partially contained events is illustrated. The
dotted histograms show the non-oscillation Monte Carlo events, and
the solid histograms show the best-fit expectations for atmospheric
$\nu^{}_\mu \to \nu^{}_\mu$ oscillations \cite{PDG}.}
\end{figure}

In 2004 the SK Collaboration carried out a careful analysis of the
$\nu^{}_\mu$ (or $\overline{\nu}^{}_\mu$) disappearance probability
as a function of the neutrino flight length $L$ over the neutrino
energy $E$, and observed a dip in the $L/E$ distribution as the
first {\it direct} evidence for atmospheric neutrino oscillations
\cite{SK04}. This dip was consistent with the prediction from the
sinusoidal flavor transition probability of neutrino oscillations,
but inconsistent with the exotic neutrino decay and neutrino
decoherence scenarios.

To directly observe the atmospheric $\nu^{}_\mu \to \nu^{}_\tau$
oscillation is quite difficult because it requires the neutrino beam
energy greater than a threshold of 3.5 GeV, such that a tau lepton
can be produced via the charged-current interaction of incident
$\nu^{}_\tau$ with the target nuclei in the detector. But the SK
data are found to be best described by neutrino oscillations that
include the $\nu^{}_\tau$ appearance in addition to the overwhelming
signature of the $\nu^{}_\mu$ disappearance. A neural network
analysis of the zenith-angle distribution of multi-GeV contained
events has recently demonstrated this observation at the $3.8
\sigma$ level \cite{SK13}.

\subsection{Accelerator neutrino oscillations}

If the observed deficit of atmospheric $\nu^{}_\mu$ and
$\overline{\nu}^{}_\mu$ events is ascribed to neutrino oscillations,
then a fraction of the accelerator-produced $\nu^{}_\mu$ and
$\overline{\nu}^{}_\mu$ events should also disappear on their way to
a remote detector. This expectation has definitely been confirmed by
two long-baseline neutrino oscillation experiments: K2K \cite{K2K}
and MINOS \cite{MINOS}. The K2K experiment was designed in such a
way that the $\nu^{}_\mu$ beam was produced at the KEK accelerator
and measured 250 km away at the SK detector in Kamioka. In
comparison, the baseline length of the MINOS experiment is 735 km,
from the source of $\nu^{}_\mu$ neutrinos at Fermilab to the far
detector in northern Minnesota. Both of them have observed a
reduction of the $\nu^{}_\mu$ flux and a distortion of the
$\nu^{}_\mu$ energy spectrum, implying $\nu^{}_\mu \to \nu^{}_\mu$
oscillations. The most striking result obtained from the atmospheric
and accelerator neutrino oscillation experiments is $\sin^2
2\theta^{}_{23} \simeq 1$ or $\theta^{}_{23} \simeq 45^\circ$, which
might hint at a special flavor structure or a certain flavor
symmetry in the neutrino sector \cite{Vissani}.

An especially important accelerator neutrino oscillation experiment is the T2K
experiment with a $\nu^{}_\mu$ beam produced from the J-PARC Main
Ring in Tokai and pointing to the SK detector at a distance of 295
km. Its main goal is to discover $\nu^{}_\mu \to \nu^{}_e$
appearance oscillations and perform a precision measurement of
$\nu^{}_\mu \to \nu^{}_\mu$ disappearance oscillations.
Since its preliminary data were first released in June 2011, the T2K
experiment has proved to be very successful in establishing the
$\nu^{}_e$ appearance out of a $\nu^{}_\mu$ beam at the $7.3 \sigma$
level and constraining the neutrino mixing parameters
$\theta^{}_{13}$, $\theta^{}_{23}$ and $\delta$ \cite{T2K}. The
point is that the leading term of $P(\nu^{}_\mu \to \nu^{}_e)$ is
sensitive to $\sin^2 2\theta^{}_{13} \sin^2\theta^{}_{23}$, and its
sub-leading term is sensitive to $\delta$ and terrestrial matter
effects \cite{Freund}. Fig. 5 shows the allowed region of $\sin^2
2\theta^{}_{13}$ changing with the CP-violating phase $\delta$ as
constrained by the T2K data \cite{T2K}, from which one can see an
unsuppressed value of $\theta^{}_{13}$ together with a preliminary
hint $\delta \sim -\pi/2$ even though the neutrino mass ordering
(i.e., the sign of $\Delta m^2_{32}$) remains undetermined.
\begin{figure}[t]
\centerline{\includegraphics[width=12.8cm]{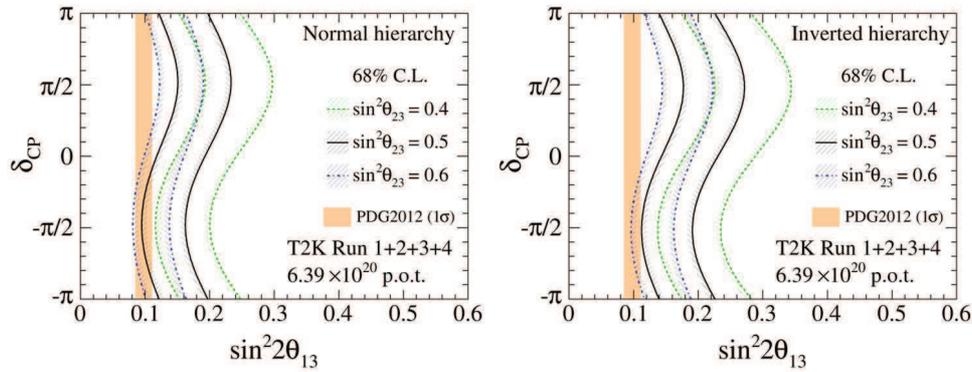}} \caption{The
allowed region of $\sin^2 2\theta^{}_{13}$ as a function of the
CP-violating phase $\delta$, constrained by the present T2K neutrino
oscillation data \cite{T2K}.}
\end{figure}

Different from the K2K, MINOS and T2K experiments, the OPERA
experiment was designed to search for the $\nu^{}_\tau$ appearance
in a $\nu^{}_\mu$ beam traveling from CERN to Gran Sasso at a
distance of 730 km. After several years of data taking, the OPERA
Collaboration reported four $\nu^{}_\tau$ candidate events in 2014.
These events are consistent with $\nu^{}_\mu \to \nu^{}_\tau$
oscillations with the $4.2 \sigma$ significance \cite{O}.

\subsection{Reactor antineutrino oscillations}

Since the first discovery of electron antineutrinos with the help of
the Savannah River reactor in 1956 \cite{RC}, reactors have been
playing an important role in neutrino physics. In particular, two of
the three neutrino mixing angles ($\theta^{}_{12}$ and
$\theta^{}_{13}$) have been measured in the KamLAND \cite{KM} and
Daya Bay \cite{DYB} reactor antineutrino oscillation experiments to
an unprecedentedly good degree of accuracy.
\begin{figure}[t]
\centerline{\includegraphics[width=9.1cm]{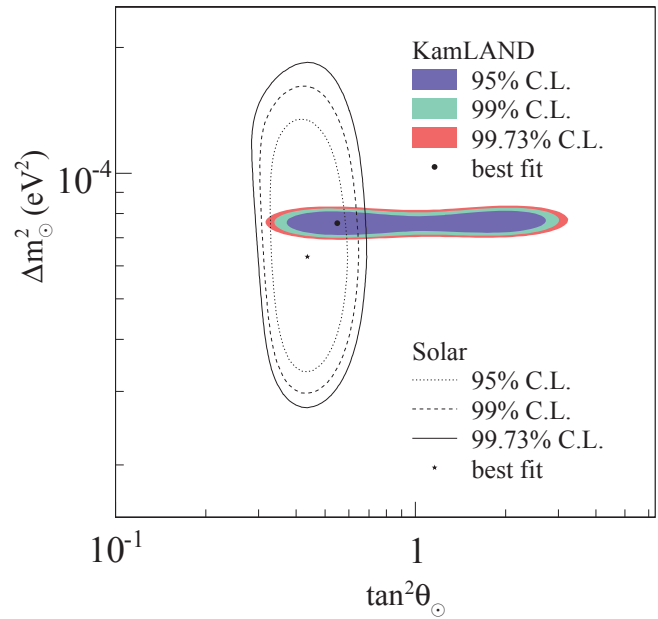}} \caption{The
allowed region for two-flavor neutrino oscillation parameters from
the KamLAND and solar neutrino experiments, where $\Delta m^2_\odot
\simeq \Delta m^2_{21}$ and $\tan^2\theta^{}_\odot \simeq
\tan^2\theta^{}_{12}$ hold \cite{KM2}.}
\end{figure}

The average baseline length of the KamLAND experiment was $L =180$
km, and hence it was sensitive to the $\Delta m^2_{21}$-driven
$\overline{\nu}^{}_e \to \overline{\nu}^{}_e$ oscillation and
allowed a terrestrial test of the large-mixing-angle (LMA) MSW
solution to the solar neutrino problem. Under CPT invariance the
KamLAND measurement \cite{KM} firmly established the LMA solution
for the first time, and pinned down the correct parameter space of
solar $\nu^{}_e \to \nu^{}_e$ oscillations constrained by the SNO
and SK experiments, as shown in Fig. 6 in the two-flavor scheme
\cite{KM2}. A striking sinusoidal behavior of $P(\overline{\nu}^{}_e
\to \overline{\nu}^{}_e)$ against $L/E$ was also demonstrated in the
KamLAND experiment \cite{KM2}.

While the CHOOZ \cite{CHOOZ} and Palo Verde \cite{PV} reactor
antineutrino experiments tried to search for the $\Delta
m^2_{31}$-driven $\overline{\nu}^{}_e \to \overline{\nu}^{}_e$
oscillations at the end of the 20th century, they found no
indication in favor of such oscillations and thus set an upper bound
on the smallest neutrino mixing angle $\theta^{}_{13}$. This
situation has been changed by the Daya Bay \cite{DYB}, RENO
\cite{RENO} and Double Chooz \cite{DC} experiments in the past few
years.

The Daya Bay experiment was designed to probe the smallest neutrino
mixing angle $\theta^{}_{13}$ with an unprecedented sensitivity
$\sin^2 2\theta^{}_{13} \sim 1\%$ by measuring the
$\Delta m^2_{31}$-driven $\overline{\nu}^{}_e \to \overline{\nu}^{}_e$
oscillation with a baseline length $L \simeq 2$ km. In this experiment
the electron antineutrino beam takes its source at the Daya Bay nuclear
power complex located in Shenzhen, as shown in Fig. 7. The eight
antineutrino detectors deployed at the near (two plus two) and far
(four) sites are all the liquid scintillator detectors. In March 2012
the Daya Bay Collaboration announced a $5.2\sigma$ discovery of
$\theta^{}_{13} \neq 0$, with $\sin^2
2\theta^{}_{13} = 0.092 \pm 0.016 ({\rm stat}) \pm 0.005 ({\rm
syst})$ (see Fig. 8 for illustration) \cite{DYB}. A similar but slightly less
significant result was later achieved in the RENO \cite{RENO} and Double
Chooz \cite{DC} reactor antineutrino experiments.
\begin{figure}[t]
\centerline{\includegraphics[width=11.5cm]{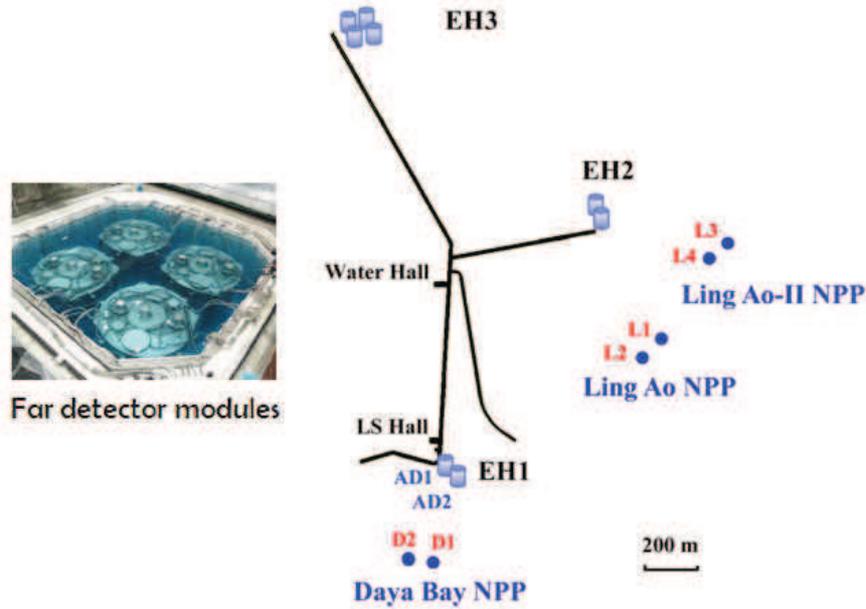}} \caption{The
layout of the Daya Bay reactor antineutrino experiment with three
pairs of reactor cores (Daya Bay, Ling Ao I and Ling Ao II). Four
detector modules are deployed at the far site, and two detector
modules are deployed at each of the two near sites \cite{DYB}.}
\end{figure}
\begin{figure}[t]
\centerline{\includegraphics[width=9.5cm]{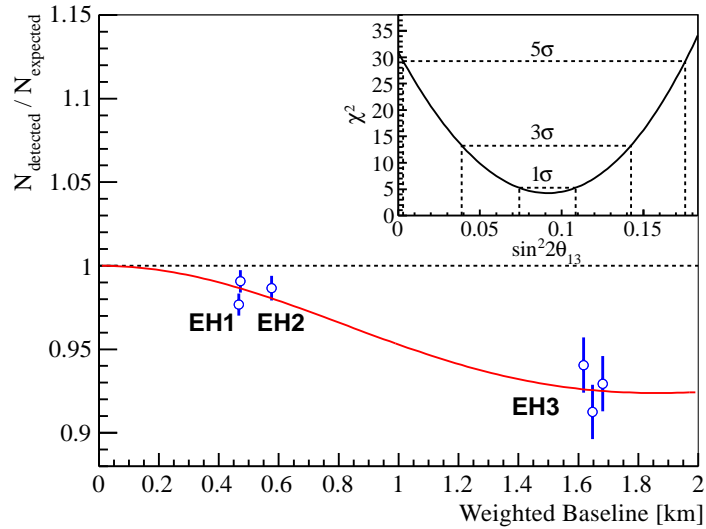}} \caption{The
survival probability of $\overline{\nu}^{}_e \to \overline{\nu}^{}_e$
oscillations observed at the near and far experimental halls (i.e., EH1,
EH2 and EH3) in the Daya Bay experiment \cite{DYB}.}
\end{figure}

The Daya Bay Collaboration has also measured the energy dependence of
$\overline{\nu}^{}_e$ disappearance and observed a nearly full
oscillation cycle against $L/E$ \cite{DYB2}.
An improved result of the oscillation amplitude $\sin^2
2\theta^{}_{13} = 0.090^{+0.008}_{-0.009}$ has recently been obtained
by using the observed $\overline{\nu}^{}_e$
rate and the observed energy spectrum in the three-flavor framework
\cite{DYB2}. The relative large value of $\theta^{}_{13}$ is very
encouraging for the next-generation precision neutrino experiments, which
aim to determine the neutrino mass ordering and probe
leptonic CP violation in the foreseeable future.

\subsection{Determination of oscillation parameters}

The aforementioned neutrino or antineutrino oscillation experiments
involve different sources, different flavors, different energies and
different baseline lengths. But the relevant experimental data can
all be explained in the scheme of three-flavor oscillations, which
depend on two independent neutrino mass-squared differences ($\Delta
m^2_{21}$, $\Delta m^2_{32}$), three flavor mixing angles
($\theta^{}_{12}$, $\theta^{}_{13}$, $\theta^{}_{23}$) and one
CP-violating phase ($\delta$). A global fit of all the available
experimental data is therefore needed in order to determine or
constrain the six oscillation parameters.

A global three-flavor analysis of current experimental
data on solar (SNO, SK,
Borexino), atmospheric (SK), accelerator (MINOS, T2K) and reactor
(KamLAND, Daya Bay, RENO) neutrino or antineutrino oscillations
has recently been done by several groups \cite{Fogli,Valle,Schwetz}.
For the sake of simplicity, here we only quote the main results
obtained by the Italian group \cite{Fogli}
\footnote{In this reference the notations $\delta m^2 \equiv m^2_2 - m^2_1$
and $\Delta m^2 \equiv m^2_3 - (m^2_1 + m^2_2)/2$ are used.
Their relations with $\Delta m^2_{21}$ and $\Delta
m^2_{31}$ are rather simple: $\Delta m^2_{21} = \delta m^2$ and
$\Delta m^2_{31} = \Delta m^2 + \delta m^2/2$.},
as listed in Table 1.
\begin{table}[t]
\tbl{The three-flavor neutrino oscillation parameters determined
or constrained from a global analysis of current experimental data
\cite{Fogli}.}
{\begin{tabular}{ccccc}
\hline
\hline
Parameter & Best fit & 1$\sigma$ range & 2$\sigma$ range & 3$\sigma$ range \\
\hline
\multicolumn{5}{c}{Normal neutrino mass ordering
$(m^{}_1 < m^{}_2 < m^{}_3$)} \\ \hline
$\Delta m^2_{21}/10^{-5} ~{\rm eV}^2$
& $7.54$  & 7.32 --- 7.80 & 7.15 --- 8.00 & 6.99 --- 8.18 \\
$\Delta m^2_{31}/10^{-3} ~ {\rm eV}^2$ & $2.47$
& 2.41 --- 2.53 & 2.34 --- 2.59 & 2.26 --- 2.65 \\
$\sin^2\theta_{12}/10^{-1}$
& $3.08$ & 2.91 --- 3.25 & 2.75 --- 3.42 & 2.59 --- 3.59 \\
$\sin^2\theta_{13}/10^{-2}$
& $2.34$ & 2.15 --- 2.54 & 1.95 --- 2.74 & 1.76 --- 2.95 \\
$\sin^2\theta_{23}/10^{-1}$
& $4.37$  & 4.14 --- 4.70 & 3.93 --- 5.52 & 3.74 --- 6.26 \\
$\delta/180^\circ$ &  $1.39$ & 1.12 --- 1.77 & 0.00 --- 0.16
$\oplus$ 0.86 --- 2.00 & 0.00 --- 2.00 \\ \hline
\multicolumn{5}{c}{Inverted neutrino mass ordering
$(m^{}_3 < m^{}_1 < m^{}_2$)} \\ \hline
$\Delta m^2_{21}/10^{-5} ~{\rm eV}^2$
& $7.54$  & 7.32 --- 7.80 & 7.15 --- 8.00 & 6.99 --- 8.18 \\
$\Delta m^2_{13}/10^{-3} ~ {\rm eV}^2$ & $2.42$
& 2.36 --- 2.48 & 2.29 --- 2.54 & 2.22 --- 2.60 \\
$\sin^2\theta_{12}/10^{-1}$
& $3.08$ & 2.91 --- 3.25 & 2.75 --- 3.42 & 2.59 --- 3.59 \\
$\sin^2\theta_{13}/10^{-2}$
& $2.40$ & 2.18 --- 2.59 & 1.98 --- 2.79 & 1.78 --- 2.98 \\
$\sin^2\theta_{23}/10^{-1}$
& $4.55$  & 4.24 --- 5.94 & 4.00 --- 6.20 & 3.80 --- 6.41 \\
$\delta/180^\circ$ &  $1.31$ & 0.98 --- 1.60 & 0.00 --- 0.02
$\oplus$ 0.70 --- 2.00 & 0.00 --- 2.00 \\ \hline\hline
\end{tabular}}
\end{table}

Table 1 shows that the output values of $\theta^{}_{13}$,
$\theta^{}_{23}$ and $\delta$ in such a global fit are sensitive to
the sign of $\Delta m^2_{31}$. That is why it is crucial to
determine the neutrino mass ordering in the upcoming neutrino
oscillation experiments. The hint $\delta \neq 0^\circ$ (or
$180^\circ$) at the $1\sigma$ level is still preliminary but quite
encouraging, because it implies a potential effect of leptonic CP
violation which is likely to show up in some long-baseline neutrino
oscillation experiments in the foreseeable future. The possibility
$\theta^{}_{23} = 45^\circ$ cannot be ruled out at the $2\sigma$
level, and thus a more precise determination of $\theta^{}_{23}$ is
required in order to resolve its octant.

It is worth pointing out that $|U^{}_{\mu i}| = |U^{}_{\tau i}|$
(for $i=1,2,3$), the so-called $\mu$-$\tau$ permutation symmetry of
the PMNS matrix $U$ itself, holds if either the conditions
$\theta^{}_{13} = 0^\circ$ and $\theta^{}_{23} = 45^\circ$ or the
conditions $\delta = 90^\circ$ (or $270^\circ$) and $\theta^{}_{23}
= 45^\circ$ are satisfied \cite{Zhou}. Now that $\theta^{}_{13} = 0^\circ$ has
definitely been excluded, it is imperative to know the values of
$\theta^{}_{23}$ and $\delta$ as accurately as possible, so as to fix
the strength of $\mu$-$\tau$ symmetry breaking associated with the
structure of $U$.

\section{Neutrino Mass Ordering and CP Violation}

The neutrino mass ordering can be explored with either reactor
electron antineutrinos or atmospheric muon neutrinos in the
``disappearance" oscillation experiments, or with accelerator muon
neutrinos in the ``appearance" oscillation experiments. Let us take
the JUNO \cite{JUNO}, PINGU \cite{PINGU} and LBNE \cite{LBNE}
experiments for example to illustrate the future prospects in this
regard.

The JUNO electron antineutrino detector is expected to be a
20-kiloton liquid-scintillator detector located in the Jiangmen city of
Guangdong province in southern China, about 53 km away from the
Yangjiang (17.4 $\rm GW^{}_{th}$) and Taishan (18.4 $\rm
GW^{}_{th}$) reactor facilities which serve as the
$\overline{\nu}^{}_e$ source. Given Eq. (5), the survival
probability of $\overline{\nu}^{}_e \to \overline{\nu}^{}_e$
oscillations can be explicitly expressed as
\begin{eqnarray}
P(\overline{\nu}^{}_e \to \overline{\nu}^{}_e) & = & 1 - \sin^2
2\theta^{}_{12} \cos^4\theta^{}_{13} \sin^2 \Delta^{}_{21} -
\frac{1}{2} \sin^2 2\theta^{}_{13} \left[ 1 - \cos\Delta^{}_{*}
\cos\Delta^{}_{21} \right. \nonumber \\ && + \left. \cos
2\theta^{}_{12} \sin\Delta^{}_{*} \sin\Delta^{}_{21} \right] \; ,
\end{eqnarray}
where $\Delta^{}_{*} \equiv \Delta^{}_{31} + \Delta^{}_{32}$. In Eq.
(6) the oscillating argument $\Delta^{}_{21}$ is unambiguous, and
the neutrino mass ordering is determined by the sign of
$\Delta^{}_{*}$ (normal: positive; inverted: negative). To
distinguish the inverted neutrino mass hierarchy from the normal
one, it is necessary to measure the $\Delta^{}_{*}$-driven
oscillations over many cycles on condition that $\Delta^{}_{21} \sim
\pi/2$ is satisfied for $L \sim 53$ km as taken in the JUNO
experiment \cite{Zhan}. Fig. 9 illustrates why this idea works.
\begin{figure}[t]
\centerline{\includegraphics[width=9.5cm]{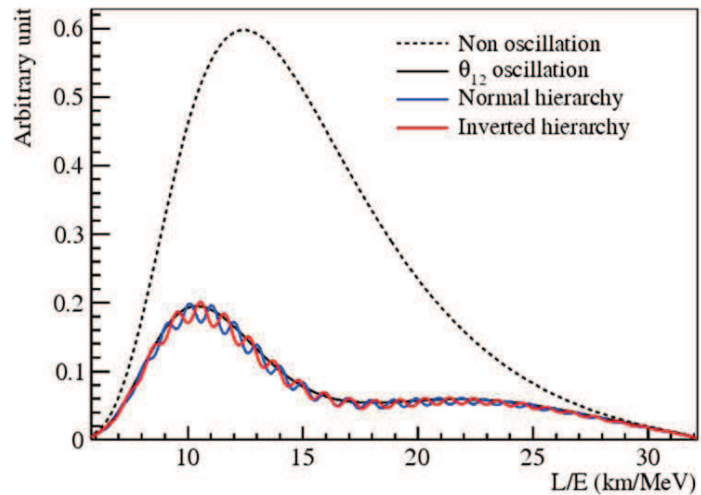}} \caption{The
reactor antineutrino spectrum changing with $L/E$ at a baseline $L
\sim 53$ km, where the blue (normal) or red (inverted) fine
structure can tell the neutrino mass hierarchy after a Fourier
transformation of the spectrum \cite{Zhan}.}
\end{figure}

Now the JUNO experiment's civil construction is underway, and its
detector assembly is planned for 2018 to 2019. Data taking will
commence in 2020, with a target of about six years of operation to
pin down the neutrino mass ordering at the $3 \sigma$ or $4 \sigma$
level \cite{JUNO}. The challenges for this experiment, which must be
met successfully, are mainly technological, such as how to improve
the scintillator light yield, attenuation length and PMT quantum
efficiency \cite{Luk}.

The PINGU experiment is a proposed low-energy infill extension of
the IceCube experiment at the South Pole \cite{PINGU}. Its design
closely follows the one used for IceCube and DeepCore. The idea is
to further infill the central DeepCore volume with 40 new strings of
60 optical modules each, so that the neutrino trigger energy
threshold can be lowered to a few GeV and thus high-quality
reconstructions for neutrino events can be achieved between 5 and 15
GeV. Such a detector geometry will be able to distinguish between
the normal and inverted neutrino mass hierarchies at the $3 \sigma$
significance with an estimated 3.5 years of data taking.

The survival probability of atmospheric muon neutrinos that reach
the PINGU detector after propagation through the Earth (i.e., from
below) depends on their beam energy $E$ and propagation length $L$.
Thanks to interactions with electrons within the Earth, a resonant
flavor conversion can happen at a specific pattern of neutrino
energies and Earth-crossing paths. This matter-induced resonant
conversion occurs only for neutrinos in the normal mass ordering or
only for antineutrinos in the inverted mass ordering, as
the behaviors of $\nu^{}_\mu \to \nu^{}_\mu$ and
$\overline{\nu}^{}_\mu \to \overline{\nu}^{}_\mu$ oscillations
depend respectively on $\Delta m^2_{31} \mp 2\sqrt{2} G^{}_{\rm F}
N^{}_e E$, where $N^{}_e$ is the number density of electrons in
matter and $E$ denotes the neutrino beam energy. The PINGU
detector is capable of discriminating the cross sections and
kinematics of neutrino and antineutrino interactions with nuclei,
so it is capable of identifying different detected event rates
which depend on different neutrino mass orderings.

Given an accelerator-driven neutrino beam, the long-baseline
oscillation experiments are also sensitive to the neutrino mass
ordering. Because of the interaction of neutrinos with terrestrial
matter as they pass through the Earth, the probability of
$\nu^{}_\mu \to \nu^{}_e$ oscillations can be approximately
expressed as \cite{Freund}
\begin{eqnarray}
P(\nu^{}_\mu \to \nu^{}_e) & \simeq & \sin^2 2\theta^{}_{13}
\sin^2\theta^{}_{23} \frac{\sin^2 \left(x -1\right) \Delta^{}_{31}}
{\left(x -1\right)^2} + \alpha \sin 2\theta^{}_{12}
\sin 2\theta^{}_{13} \sin 2\theta^{}_{23} \nonumber \\
&& \times \cos\left(\Delta^{}_{31}
+ \delta\right) \frac{\sin x \Delta^{}_{31}
\sin\left(x -1\right) \Delta^{}_{31}}{x \left(x -1\right)} \nonumber \\
&& + \alpha^2 \sin^2 2\theta^{}_{12} \cos^2\theta^{}_{23}
\frac{\sin^2 x \Delta^{}_{31}}{x^2} \; ,
\end{eqnarray}
where $x\equiv 2\sqrt{2} G^{}_{\rm F} N^{}_e E/\Delta m^2_{31}$ and
$\alpha \equiv \Delta m^2_{21}/\Delta m^2_{31}$. One may easily obtain
the expression of $P(\overline{\nu}^{}_\mu \to \overline{\nu}^{}_e)$
from Eq. (7) with the replacements $\delta \to -\delta$ and
$x \to -x$. So the sign of
$\Delta m^2_{31}$ affects the behaviors of neutrino oscillations via
the signs of $x$ and $\alpha$. That is why the matter-induced
resonant conversion can only occur for neutrinos in the normal mass
hierarchy ($x >0$) or for antineutrinos in the inverted mass hierarchy
($x <0$), similar to the case of atmospheric neutrino or antineutrino
oscillations. In practice the baseline length $L$ of an experiment
is crucial for its sensitivity to the mass hierarchy. The LBNE
experiment \cite{LBNE} with $L \simeq 1300$ km is therefore expected
to be more promising than the T2K experiment \cite{T2K} with $L
\simeq 295$ km and the NO$\nu$A experiment \cite{NOVA} with $L
\simeq 810$ km in this respect. But the undetermined CP-violating
phase $\delta$ may in general give rise to some uncertainties
associated with a determination of the neutrino mass hierarchy in
the long-baseline experiments. In particular, a careful analysis
shows that the mass hierarchy sensitivity is most optimistic (or
pessimistic) for $\delta \simeq -\pi/2$ in the normal (or inverted)
hierarchy case, or for $\delta \simeq +\pi/2$ in the inverted (or
normal) hierarchy case \cite{LBNE}. Regardless of possible values of
$\delta$, LBNE in combination with T2K and NO$\nu$A promises to
resolve the neutrino mass hierarchy with a significance of more than
$3\sigma$ by 2030 \cite{Luk}.

In addition, the proposed Hyper-Kamiokande (HK) detector will be a
next-generation underground water Cherenkov detector serving as the
far detector of the 295 km-baseline neutrino oscillation experiment for
the J-PARC neutrino beam \cite{HK}. It is expected to be ten times
larger than the SK detector and capable of probing the neutrino
mass ordering, resolving the octant of the largest flavor mixing angle
$\theta^{}_{23}$ and observing leptonic CP violation as well as proton
decays and extraterrestrial neutrinos from distant astrophysical sources.

CP violation in the lepton sector may have far-reaching impacts on
our understanding of the origin of matter-antimatter asymmetries
at both microscales and macroscales. The LBNE and HK
experiments, together with other next-generation long-baseline neutrino
oscillation experiments, are aiming at a determination of the
CP-violating phase $\delta$. The latter can be extracted from comparing
between the probabilities of $\nu^{}_\mu \to \nu^{}_e$ and
$\overline{\nu}^{}_\mu \to \overline{\nu}^{}_e$ oscillations, but it
is in general contaminated by terrestrial matter effects.
In the leading-order approximation,
\begin{eqnarray}
{\cal A}^{}_{\rm CP} \equiv \frac{P(\nu^{}_\mu \to \nu^{}_e) -
P(\overline{\nu}^{}_\mu \to \overline{\nu}^{}_e)}
{P(\nu^{}_\mu \to \nu^{}_e) +
P(\overline{\nu}^{}_\mu \to \overline{\nu}^{}_e)}
\simeq -\frac{\sin 2\theta^{}_{12} \sin\delta}
{\sin\theta^{}_{13} \tan\theta^{}_{23}} \Delta^{}_{21} +
{\rm matter ~ effects} \; ,
\end{eqnarray}
where the term of matter effects should more or less be correlated
with the neutrino mass ordering. To lower the matter contamination,
one may therefore consider a low-energy neutrino (or antineutrino) beam
with a much shorter baseline length \cite{Low}. A proposal of this
kind is the MOMENT project with a neutrino beam energy
$E \sim 300$ MeV and a baseline length $L \sim 120$ km \cite{MOMENT},
towards probing leptonic CP violation before a more powerful
neutrino factory is built.

\section{Two Non-oscillation Aspects}

\subsection{Neutrinoless double-beta decays}

Soon after Fermi developed an effective beta decay theory
\cite{Fermi}, Maria Goeppert-Mayer pointed out that certain
even-even nuclei should have a chance to decay into the second
nearest neighbors via two simultaneous beta decays \cite{Mayer}:
$(A, Z) \to (A, Z+2) + 2 e^- + 2 \overline{\nu}^{}_e$, where the
kinematic conditions $m (A, Z) > m (A, Z+2)$ and $m (A, Z) < m (A,
Z+1)$ must be satisfied. In 1939 Wendell Furry further pointed out
that the $0\nu\beta\beta$ decays $(A, Z)
\to (A, Z+2) + 2 e^-$ could happen via an exchange of the {\it
virtual} neutrinos between two associated beta decays \cite{Furry},
provided the neutrinos are massive and have the Majorana nature
\cite{Majorana}. If such a $0\nu\beta\beta$ process is measured,
does it definitely imply the existence of a Majorana mass term for
neutrinos? The answer is affirmative according to the
Schechter-Valle theorem \cite{SV}, no matter whether there are new
physics contributions to the $0\nu\beta\beta$ decays. Hence the
$0\nu\beta\beta$ transitions can serve for an experimentally
feasible probe towards identifying the Majorana nature of massive
neutrinos at low energies.

The half-life of a
$0\nu\beta\beta$-decaying nuclide can be expressed as follows:
\begin{eqnarray}
T^{0\nu}_{1/2} = \left(G^{0\nu}\right)^{-1} \left|M^{0\nu}\right|^{-2}
\left|\langle m\rangle^{}_{ee}\right|^{-2} \; , ~~~~~~ \langle
m\rangle^{}_{ee} \equiv \sum^{}_{i} \left( m^{}_i U^2_{e i} \right)
\; ,
\end{eqnarray}
where $G^{0\nu}$ is the phase-space factor, $M^{0\nu}$ stands for
the relevant nuclear matrix element, and $\langle m\rangle^{}_{ee}$
denotes the effective Majorana neutrino mass in the absence of
new physics contributions. Among them, the
calculation of $|M^{0\nu}|$ relies on the chosen nuclear models
which are only able to approximately describe the many-body
interactions of nucleons in nuclei, and thus it involves the largest
theoretical uncertainty (e.g., a factor of two or three for some
typical nuclei) \cite{Giunti}. This causes quite a big uncertainty
associated with the determination of $|\langle
m\rangle^{}_{ee}|$.

So far no convincing evidence for an occurrence of the
$0\nu\beta\beta$ decay has been established, although a lot of
experimental efforts have been made in the past few decades. Such an
experiment is designed to observe the two electrons emitted in a
given $0\nu\beta\beta$ decay, and its signature is based on the
fact that the sum of the energies of the two emitted electrons is
equal to the $Q$-value of this process. In contrast, the energy
spectrum of the two emitted electrons in a normal double-beta decay must
be continuous. At present the strongest upper bound on the effective
mass term $|\langle m\rangle^{}_{ee}|$ can be set by the
$^{76}_{32}{\rm Ge} \to ~^{76}_{34}{\rm
Se} + 2 e^-$ and $^{136}_{~54}{\rm Xe} \to ~^{136}_{~56}{\rm Ba} + 2
e^-$ experiments \cite{Giunti}. In particular, the GERDA \cite{GERDA},
EXO-200 \cite{EXO} and KamLAND-Zen \cite{KZ} experiments have obtained
$T^{0\nu}_{1/2} > 2.1 \times 10^{25} ~{\rm yr}$,
$1.1 \times 10^{25} ~{\rm yr}$
and $1.9 \times 10^{25} ~{\rm yr}$
at the $90\%$ confidence level, respectively.
These results lead to the constraints
$|\langle m\rangle^{}_{ee}| < 0.22$---$0.64$ eV,
$0.2$---$0.69$ eV and $0.15$---$0.52$ eV at the same confidence
level, respectively, after the relevant
uncertainties of nuclear matrix elements are taken into
account \cite{Giunti}.
\begin{figure}[t]
\centerline{\includegraphics[width=7.5cm]{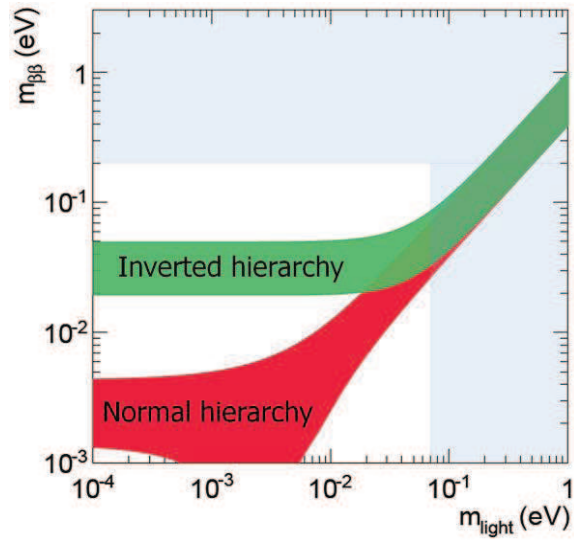}} \caption{The
effective Majorana neutrino mass $m^{}_{\beta\beta} \equiv |\langle
m\rangle^{}_{ee}|$ as a function of the lightest neutrino mass
$m^{}_{\rm light} \equiv m^{}_1$ (normal hierarchy, red band) or
$m^{}_3$ (inverted hierarchy, green band) \cite{GC}. Here the
horizontally-excluded region comes from the $0\nu\beta\beta$
experiments \cite{GERDA,EXO,KZ}, and the vertically-excluded region
is due to the cosmological bound \cite{Planck}.}
\end{figure}

The expected magnitude of $|\langle m\rangle^{}_{ee}|$ in the
standard three-flavor case is illustrated in Fig. 10, where current
neutrino oscillation data have been input and arbitrary values of
the CP-violating phases have been taken \cite{GC}. It is clear that
the inverted neutrino mass ordering or
a near neutrino mass degeneracy may allow $|\langle
m\rangle^{}_{ee}| \geq 0.01$ eV, which should be accessible in the
next-generation $0\nu\beta\beta$-decay experiments. If the neutrino
mass spectrum is normal and hierarchical, however, there will be
little prospect of observing any $0\nu\beta\beta$ decays in the
foreseeable future, simply because of $|\langle m\rangle^{}_{ee}|
\sim {\cal O}(10^{-3})$ eV in this unfortunate case.

\subsection{The absolute neutrino mass scale}

Since the flavor oscillations of massive neutrinos are only
sensitive to the neutrino mass-squared differences, a determination
of the absolute neutrino mass scale has to rely on some
non-oscillation experiments. Searching for the $0\nu\beta\beta$
decay is one of the feasible ways for this purpose if massive
neutrinos are the Majorana particles, because the magnitude of its
effective mass term $\langle m\rangle^{}_{ee}$ is associated
with $m^{}_i$ as shown in Eq. (9) and Fig. 10. Another way is to
detect the beta decays, such as $^3_1 {\rm H} \to ~ ^3_2 {\rm He} + e^-
+ \overline{\nu}^{}_e$, whose effective neutrino mass term
$\langle m\rangle^{}_e$ is defined via
\begin{eqnarray}
\left(\langle m\rangle^{}_e\right)^2 \equiv \sum_i
\left(m^2_i |U^{}_{e i}|^2 \right) \; .
\end{eqnarray}
The most promising experiment of this kind
is the KATRIN experiment \cite{KATRIN}, which may hopefully probe
$\langle m\rangle^{}_e$ with a sensitivity of about $0.2$ eV in
the near future. But up to now only
$\langle m\rangle^{}_e < 2.05$ eV has been obtained at the $95\%$
confidence level from the Troitzk beta-decay experiment \cite{Beta}.

Furthermore, one may get useful information on the mass scale of
light neutrinos from cosmology. Based on the
standard $\Lambda$CDM model, a global analysis
of current cosmological data (especially those on the cosmic
microwave background (CMB) radiation and large-scale structure (LSS)
formation) can provide us with the most powerful
sensitivity to the sum of light neutrino masses via the relation
\begin{eqnarray}
\Omega^{}_\nu h^2 = \frac{1}{\rm 93 ~ eV} \Sigma^{}_\nu \; , ~~~~~~
\Sigma^{}_\nu  \equiv \sum_i m^{}_i \; ,
\end{eqnarray}
in which $\Omega^{}_\nu$ denotes the light neutrino contribution to
today's energy density of the Universe, and $h$ is the Hubble constant.
For example, $\Sigma^{}_\nu < 0.23$ eV has recently been reported by
the Planck Collaboration at the $95\%$ confidence level \cite{Planck}.
If a combination of the next-generation CMB and LSS measurements
can reach a sensitivity of about $0.02$ eV for the sum of three
neutrino masses \cite{Abazajian}, then it will be possible to
determine the absolute neutrino mass scale via a definite
determination of $\Sigma^{}_\nu$ even though the neutrino mass ordering
is normal.

Note that it is also possible to determine or constrain the absolute
neutrino mass scale $m^{}_\nu$ through the study of kinematic effects
of supernova neutrinos, because their flight time from a supernova's core
to a terrestrial detector will be more or less delayed as compared
with the massless particles \cite{Z}.
A careful analysis of the $\overline{\nu}^{}_e$
events from the Supernova 1987A explosion led us to an upper bound
of about $6$ eV on $m^{}_\nu$ \cite{SN}.
The prospects of this astrophysical approach depend on the emergence
of new neutrino detectors or the existence of antineutrino pulses
in the first instants of a supernova explosion \cite{V}.
Given the JUNO liquid scintillator detector as an example,
$m^{}_\nu < 0.83 \pm 0.24$ eV is expected to be
achievable at the $95\%$ confidence level for a typical galactic
supernova at a distance of $10$ kpc from the Earth \cite{Lu}.

\section{Summary and Outlook}

Since 1998, quite a lot of significant breakthroughs have been
made in experimental neutrino physics. On the one hand, the
exciting phenomena of atmospheric, solar, reactor and accelerator
neutrino or antineutrino oscillations
have all been observed, and the oscillation parameters
$\Delta m^2_{21}$, $|\Delta m^2_{31}|$, $\theta^{}_{12}$,
$\theta^{}_{13}$ and $\theta^{}_{23}$ have been determined to
an impressive degree of accuracy. On the other hand, the
geo-antineutrino events and extraterrestrial PeV neutrino events
have been observed, and the sensitivities to neutrino masses
in the beta decays, $0\nu\beta\beta$ decays and cosmology have
been improved to a great extent. Furthermore, a lot of theoretical efforts
have also been made towards understanding the origin of tiny neutrino
masses and the flavor structure behind the observed neutrino
mixing pattern, and towards studying
possible implications of massive neutrinos on
the cosmological matter-antimatter asymmetry, warm dark matter and
many violent astrophysical processes \cite{XZ,Altarelli}.
All these have demonstrated neutrino
physics to be one of the most important frontiers of
particle physics, astrophysics and cosmology.

But a number of fundamental questions about massive neutrinos remain
open. The burning ones include how small the absolute
neutrino mass scale is, whether the neutrino mass spectrum is
normal or inverted, whether massive neutrinos are the Majorana particles,
how large the CP-violating phase $\delta$ is, which octant the largest
flavor mixing angle $\theta^{}_{23}$ belongs to, whether there
are light and (or) heavy sterile neutrinos, what the role of
neutrinos is in dark matter, whether the observed matter-antimatter
asymmetry of the Universe is related to CP violation
in neutrino oscillations, etc. Motivated by so many questions,
we are trying to discover a new physics world with the help of
massive neutrinos in the coming decades.

We would like to thank Luciano Maiani and Gigi Rolandi for inviting
us to contribute to this book. We are also grateful to Yu-Feng Li,
Jue Zhang, Zhen-hua Zhao, Shun Zhou and Ye-Ling Zhou for their helpful
comments on this essay. This work is
supported in part by the National Natural Science
Foundation of China under grant No. 11135009 and
11390380; by the National Basic Research Program of China under
grant No. 2013CB834300; by the Strategic Priority Research
Program of the Chinese Academy of Sciences (CAS) under grant No.
XDA10000000; and by the CAS Center for Excellence in Particle Physics.


\end{document}